\documentstyle[aps,prb,multicol,epsf]{revtex}
\begin{document}
\draft 

\title{Ultrasonic attenuation in clean
	d-wave superconductors.}
\author{I. Vekhter$^1$,  E. J. Nicol$^1$, J. P. Carbotte$^2$}
\address{${}^1$Department of Physics, University of Guelph, Guelph, 
	Ontario N1G 2W1, Canada\\
	${}^2$Department of Physics and Astronomy, McMaster University,
	Hamilton, Ontario  L8S 4M1,  Canada}
\date{\today}
\maketitle
\begin{abstract}
We consider the attenuation of longitudinal ultrasonic waves in a
clean two-dimensional d-wave superconductor. We show that
the attenuation coefficient is linear in temperature at low temperatures for 
{\it all} in-plane directions of the 
propagation of the ultrasound, and that the 
coefficient of the linear term can be used to determine the parameters 
crucial for  
the low temperature transport in these compounds.
\end{abstract}

\begin{multicols}{2}
Much of the novel physics associated with the superconducting 
state of the high-$T_c$ cuprates
is due to non-trivial angular dependence
of the energy gap, $\Delta_{\bf k}$, at the Fermi surface.
While in conventional, s-wave, superconductors
the energy gap is finite for all quasiparticle excitations, 
it is believed that in many cuprates it has the angular structure of 
a $d_{x^2-y^2}$ state, with
lines of nodes, which leads to a gapless excitation spectrum 
along certain directions in 
momentum space. 
Consequently the temperature dependence
of the thermodynamic and transport coefficients 
in the high-$T_c$ materials is qualitatively different from that of their 
s-wave counterparts: at low temperatures
the behavior of the coefficients measuring an average of 
a particular quantity over the Fermi surface is governed
by the low-energy excitations near the nodes of the gap at the Fermi surface, 
and the resulting temperature dependence is given by power laws rather than
by exponentially decaying functions with an activation 
barrier\cite{somereview}. While the power laws are entirely 
determined by the dimension of the manifold  where the gap vanishes,
in this case 
by the existence of lines of nodes, 
the coefficients can be used to determine parameters of the superconducting
materials and test the agreement with specific models.

The ratio of the Fermi velocity at the nodes, $v_f$,
and the
velocity $v_2$ associated with the growth of the superconducting gap
at the Fermi surface 
$\Delta({\bf p}_f)={\bf v}_2({\bf p}_f-{\bf p}_f^{(n)})$ near the node
is a particularly important parameter:
both the universal limit of transport coefficients\cite{lee1}
and the temperature dependence of the penetration depth\cite{lee}
depend solely on $v_2/v_f$. The numerical value of this ratio
has not been clearly determined yet; Lee and Wen\cite{lee} 
obtain $v_f/v_2=6.8$ from the analysis of the penetration depth
data\cite{bonn1}, while the measurements of the thermal conductivity\cite{may}
 yield   $v_f/v_2=13.6$. It is therefore important to have additional
ways of experimentally determinining this parameter.

Here we analyze the electronic attenuation of the 
longitudinal ultrasound waves in a d-wave superconductor
and argue that it can be used to
measure the ratio $v_f/v_2$. 
We consider the clean case, $ql\gg 1$, 
where ${\bf q}$ is the wave vector of the sound wave, and $l$ 
is the electron mean free path, this implies that
for a linear frequency of 150MHz, taking the sound velocity
$v_s=4\times10^5$cm/sec \cite{bhatt} we need $l\ge 4\mu m$.
In previously studied optimally doped YBCO
samples the mean free path was determined
to be $l\simeq 0.6\mu m$ at $T\simeq 20$K \cite{ong},
which is closer to the hydrodynamic limit, $ql \ll 1$, where
the longitudinal
and transverse ultrasonic attenuation 
have been analyzed\cite{maki,moreno}. 
However, the ultra high purity crystals recently grown in
BaZrO$_3$ crucibles \cite{bonn2} are
about an order of magnitude cleaner, and 
experiments demonstrate that at low temperatures
$l\simeq 4-5\mu m$\cite{bonn3}, 
indicating that the limit $ql\geq 1$ is 
likely to be achieved. Experimentally the clean
limit is manifested in the attenuation rate which is linear, rather than
quadratic, in the frequency of the ultrasound waves, 
and by an angular dependence of the longitudinal
attenuation rate, described below, which differs from that
predicted in the dirty limit \cite{maki}.

Let us first consider the problem qualitatively.
In the clean limit $\alpha_s$ is determined
by a scattering rate of a phonon  by the quasiparticles. Apart from the 
matrix element of the electron-phonon interaction this scattering rate depends
 solely on the phase space available for the phonon to decay. 
As the energy and the wave vector of the sound wave are 
small on the scale of electronic excitations, 
the relevant scattering processes are 
those with negligible energy and momentum transfer, and are given by the
condition ${\bf v}_g\cdot {\bf q}=0$, 
where ${\bf v}_g$ is the group velocity of
the quasiparticles.
In a normal metal
the scattering is 
restricted to a belt on the Fermi surface where 
the Fermi velocity, ${\bf v}_f$,
is perpendicular to {\bf q}.
The opening of an s-wave gap preserves the direction of the
group velocity of the quasiparticles but thermally suppresses 
the occupancy of the states near the Fermi surface resulting  
in the exponential
suppression of the 
scattering \cite{morse}.
For an anisotropic s-wave superconductor
the attenuation 
at temperatures lower than the minimal gap still decays exponentially, 
albeit with an exponent which depends on the direction\cite{pokr}.

Recently this approach has been transferred to a d-wave superconductor
\cite{kostur,swihart}. Assuming that the quasiparticles contributing 
to the attenuation are still located at the points of the
Fermi surface where ${\bf v}_f\cdot{\bf q}=0$, 
the authors of Refs.\onlinecite{kostur,swihart}
noticed that the scattering processes sample 
the local gap in the direction
perpendicular to {\bf q}, 
and predicted the
fourfold oscillations of $\alpha_s$ as a function of the 
angle 
of propagation in the plane, $\theta$. In particular, since the gap vanishes 
along the nodal direction, the attenuation in that direction was predicted 
to be temperature independent and equal to that at $T_c$, while the 
attenuation in
other directions was determined to decay exponentially 
with the activation energy given by the local gap $\Delta(\theta\pm\pi/2)$.

However for an anisotropic order parameter not only the magnitude but also the
 direction of the group velocity changes with the opening of the 
energy gap. This change has important consequences for 
superconductors with gap nodes, especially
for compounds with a relatively 
large value of the superconducting gap amplitude, $\Delta_0$,
as a fraction of the Fermi energy  $\epsilon_f$, and at temperatures
below $\Delta_0^2/\epsilon_f$. This was first noted 
in the context of 
the heavy fermion superconductors
by Coppersmith and Klemm\cite{sue}, who predicted a maximum
in the ultrasonic attenuation for 
certain directions of the propagation of the ultrasound, 
and a power law behavior for low temperatures. 
Heavy fermion materials are 
not in the clean limit, and $T_c\sim 1$K 
made the desired  regime impossible to achieve experimentally.
In the cuprates, on the other hand,
a high transition temperature and a large ratio of 
$\Delta_0/T_c\sim$2.5-4 bring the gap amplitude to about 
10-30\% of the Fermi energy, so that the electronic contribution to
the attenuation rate is measured at $T\ll\Delta_0^2/\epsilon_f$,
where the theoretical analysis for a d-wave superconductor is still lacking.

We first proceed with the qualitative analysis.
Below $T_c$
the excitation spectrum
has the form $E({\bf k})=\sqrt{\zeta_{\bf k}^2+\Delta_{\bf k}^2}$, 
where $\zeta_{\bf k}$ is the quasiparticle energy in the normal state 
with respect to $\epsilon_f$, so that the direction of the group velocity,
given by a tangential line to a $E({\bf k})=const$ contour, 
does not coincide with the direction of the Fermi velocity at the same point.
For $T\ll T_c$ only the contours with 
$E\ll\Delta_0$, near the gap nodes, are important for
scattering. 
Then, as shown in Fig. 1a,  the contour of $E=const$ is approximately 
an ellipse $\zeta(\phi)=\pm \sqrt{E^2-\Delta^{'2}(\phi-\phi_n)^2}$,
where $\phi$ parameterizes the Fermi surface, and 
$\Delta'$ is the angular derivative of the gap at the node, and
for {\it any} wave vector ${\bf q}$ there
are two quasiparticle scattering processes
which contribute to the attenuation. 
Since $\alpha_s$ is proportional to the phase space volume 
available for scattering 
an immediate conclusion is that the attenuation
is linear in $T$ at low temperatures in {\it all}
the directions except normal
to the nodes.

A second observation concerns the role of the relatively 
large ratio $\Delta_0/\epsilon_f$. 
This can be accounted for by introducing the curvature of the Fermi surface,
as shown in Fig.1b. Then 
the group velocity of
the quasiparticles with energies $E\ll \Delta_0$ or $E\gg \Delta_0$
does not change significantly.  For energies
$E\sim \Delta_0$, however, the competition between the curvature
of the contour $E=const$ and that of the underlying Fermi surface
opens up 
additional phase space for scattering of the ultrasound propagating  
\begin{figure}
\epsfxsize=3.5in
\epsfbox{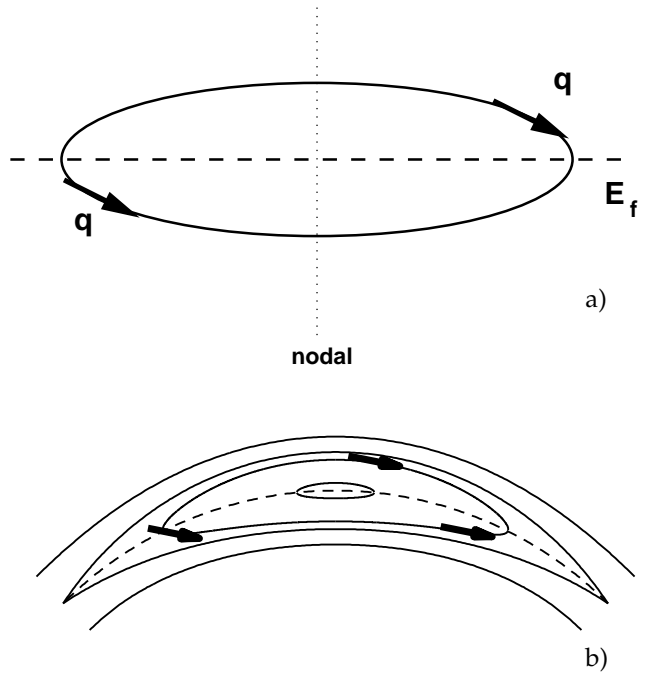}
\caption{\narrowtext a) A typical contour of constant energy 
near a node of the order parameter and the
scattering processes contributing to the attenuation for an arbitrary {\bf q};
b) Contours of constant energy $E=a\Delta_0$ for $a=0.2;0.8;1.0;1.4$
with $\Delta_0/\epsilon_f=0.1$,
and the scattering processes for {\bf q} nearly normal to the node. Arrows
{\it are} parallel.}
\label{fig:eflat}
\end{figure}\noindent
within a narrow window of angles nearly perpendicular 
to the node; 
as seen in Fig.1b 
three rather than two scattering processes become possible. 
For a d-wave gap the curvatures become comparable
for $E\sim\Delta^{'2}/2 \epsilon_f\simeq 2\Delta_0^2/\epsilon_f$, 
which implies
that the additional scattering is most effective for
$T\simeq 2\Delta_0^2/\epsilon_f$, a relatively high temperature, 
resulting in an attenuation coefficient larger than that at $T_c$.
For $T\leq T_c$ 
the additional scattering is effective,
while at lower temperatures the occupancy of these states is thermally 
suppressed and they do not contribute to the attenuation.
We therefore expect a maximum in $\alpha_s$, 
below the superconducting transition
followed by a transition to linear, in $T$, behavior at lower temperature,
$T^*$. 
The position of the maximum, as well as crossover temperature $T^*$
scale with $\Delta_0^2/\epsilon_f$,  while the angular window 
within which it exists scales as $\Delta_0^2/\epsilon_f^2$.

This 
analysis implies that the main conclusions of Ref.\onlinecite{sue}
are qualitatively applicable to d-wave superconductors. We now investigate 
the issue in more detail focusing on the low-temperature regime.
Within a standard BCS-like theory the ultrasonic attenuation is given by
\cite{sue,schr}
\end{multicols}
\widetext
\hrule width 3.45in \hfill 
\begin{equation}
\alpha_s(T, {\bf q})= A{\Omega\over 4T}
\int d{\bf k} {1\over \cosh^2(E_{\bf k}/2T)}
\Biggl[ 1+
{\zeta_{\bf k}\zeta_{{\bf k}+{\bf q}}-\Delta_{\bf k}\Delta_{{\bf k}+{\bf q}}
\over E_{\bf k}E_{{\bf k}+{\bf q}}}\Biggr]
\delta(E_{\bf k}-E_{{\bf k}+{\bf q}}),
\end{equation}
\hspace*{3.55in}\hrulefill
\begin{multicols}{2}\noindent
\begin{figure}
\epsfxsize=3.5 in
\epsfbox{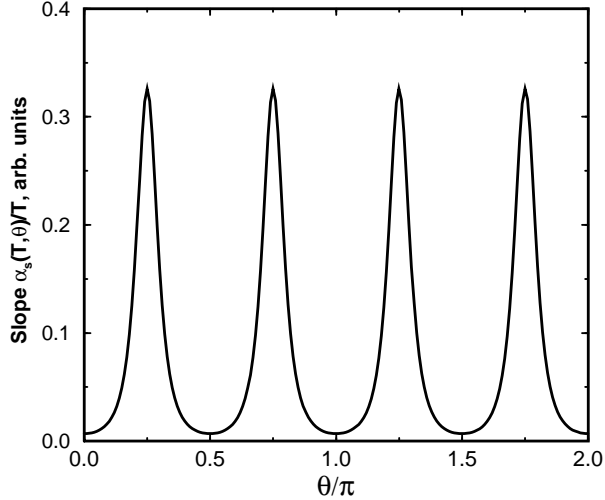}
\label{fig:slope}
\narrowtext
\caption{Angular dependence of the 
low temperature slope of the attenuation 
for $v_2/v_f=0.2$.}
\end{figure}\noindent
where the constant $A$ depends on the 
scattering matrix element and the frequency of the ultrasound, 
$\Omega$,  
and the integration is over the Brillouin zone.
Expanding in {\bf q} 
and taking into account that the angular average of the gap derivative
vanishes, 
we obtain
\begin{equation}
\alpha_s(T, {\bf q})=A {\Omega\over 2T}
\int d{\bf k} {1\over \cosh^2(E_{\bf k}/2T)}
{\zeta_{\bf k}^2\over E_{\bf k}^2}
\delta({\bf v}_g\cdot {\bf q}),
\end{equation}
where the group velocity ${\bf v}_g$ is given by
\begin{equation}
{\bf v}_g({\bf k})=\nabla_{\bf k} E_{\bf k}=
{1\over E_{\bf k}}\Bigl(\zeta_{\bf k} 
{\bf v}_f 
+ \Delta_{\bf k} {\partial\Delta_{\bf k}\over \partial{\bf k}}\Bigr).
\label{main}
\end{equation}
We now consider a model two-dimensional d-wave superconductor with the
wave vector {\bf q} in the plane at an angle $\theta$ from the x-axis.
For {\bf q} not {\it exactly} in the nodal or anti-nodal direction
${\bf v}_f\cdot {\bf q}=0$ does not satisfy the condition imposed by the delta 
function; we exclude these two specific cases from the
analysis below, although we allow $\theta$ to be arbitrarily close to these 
directions,
so that the restriction has no bearing on the generality of the results.
In the low-temperature regime
$T\ll T_c$ 
\end{multicols}
\widetext
\hrule width 3.45 in
\begin{equation}
\label{slope}
{\alpha_s(T,\theta)\over\alpha(T_c)}\approx
\ln 2{v_2\over v_f}{T\over \epsilon_f}
\Biggl[{\sin^2(\theta-\pi/4)\over
\bigl[\cos^2(\theta-\pi/4)+(v_2/v_f)^2\sin^2(\theta-\pi/4)\bigr]^{3/2}}
+
{\cos^2(\theta-\pi/4)\over
\bigl[\sin^2(\theta-\pi/4)+(v_2/v_f)^2\cos^2(\theta-\pi/4)\bigr]^{3/2}}
\Biggr].
\end{equation}
\hspace*{3.45 in}\hrulefill
\begin{multicols}{2}\noindent
Even though there is a normalization constant $\alpha(T_c)$ here, 
the ratio of the attenuation coefficients at different angles is a function
of $v_2/v_f$ only 
and therefore provides a direct measurement of this important parameter. 
Notice also that the slope vanishes as $v_2/v_f\rightarrow 0$. 
Eq.(\ref{slope})
represents the main result of this Communication, and in Fig.2
we show the slope as a function of the direction in the plane.

To obtain an explicit temperature dependence and compare
our results with those of Ref.\onlinecite{sue} in more detail
we now consider a 
superconductor with a cylindrical Fermi surface. 
Then the attenuation is given by
\begin{equation}
\label{integral}
\alpha_s(T, \theta)={A'\over 2T}{\Omega\over v_f q}
\int_0^{2\pi}{d\phi \over |\cos(\theta-\phi)|}
{\zeta_0^2(\phi)\over E(\phi)}{\cosh^{-2}\Bigl({E(\phi)\over2T}\Bigr)},
\end{equation}
where 
$\zeta_0(\phi)={\Delta_0^2}\sin 4\phi\tan(\theta-\phi)/2\epsilon_f$,
and 
$E(\phi)= \sqrt{\zeta_0^2(\phi)+\Delta^2(\phi)}$.
Notice that the contribution to the integral
of the regions $\phi\approx \theta\pm \pi/2$, 
where ${\bf v}_f\cdot {\bf q}=0$ and 
$\cos(\theta-\phi)$ tends to zero,
is suppressed exponentially by the thermal function
as $\tan(\theta-\phi)$, and consequently $E(\phi)$,  becomes large.

In Fig.3  we present the results 
for the ultrasonic attenuation normalized by its value 
at the transition temperature for near-nodal (as the nodes 
are at $m\pi/4$, the direction normal to a  node coincides 
with the nodal direction) and anti-nodal orientation of the
wave vector {\bf q} 
\begin{figure}
\epsfxsize=3.5 in
\epsfbox{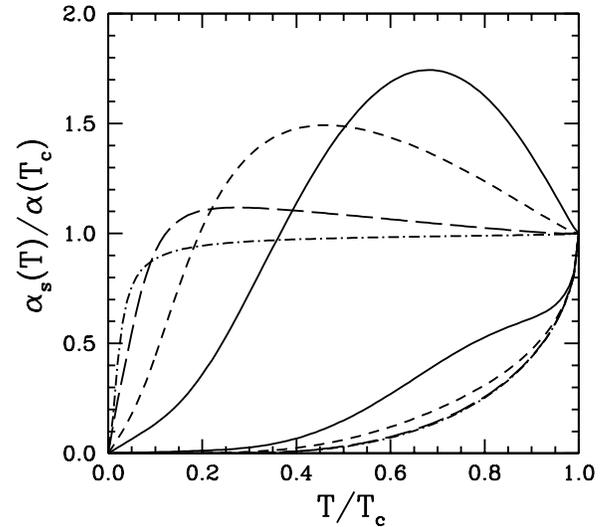}
\label{fig:temp}
\narrowtext
\caption{ Temperature dependence of the attenuation coefficient in 
the near-nodal ($\theta=0.78$, upper curves) and anti-nodal 
($\theta=1.57$, lower curves) directions for ratios of 
$\Delta_0/\epsilon_f$ of 0.25 (solid line), 0.1 (dashed line), 
0.025 (long-dashed line), and 0.0025 (dot-dashed line).}
\end{figure}\noindent
for different values of the ratio of 
$\Delta_0/\epsilon_f$, which are very similar in general 
features to those obtained numerically in Ref.\onlinecite{sue}.
In the numerical work 
we have used the BCS ratio $\Delta_0=2.14 T_c$, a larger coefficient
would push both the maximum and the onset of linear
behavior towards higher temperatures.
For the near nodal direction there is a clearly defined maximum of the 
attenuation at high temperatures, the position of the maximum scales with 
$\Delta_0^2/\epsilon_f$ as predicted above. 
The attenuation is linear at low temperatures. 
For the anti-nodal direction the decay of 
$\alpha_s(T)$ is qualitatively close to the exponential 
behavior, 
although the dependence on the ratio  $\Delta_0/\epsilon_f$ is clearly seen, 
and the shoulder on the curve for the largest value of the ratio 
indicates that
the additional phase space for scattering has become available even for
{\bf q} near the anti-node. For both angles the results for the 
smallest ratio of $\Delta_0/\epsilon_f=0.0025$ are 
indistinguishable on the scale of this graph from the 
generalized BCS prediction \cite{pokr,kostur,swihart} that 
$\alpha_s(T,\theta)/\alpha_n=f(\Delta(\theta+\pi/2))$, 
where $f$ is the Fermi function. 
We also obtain that for  a fixed value of the ratio $\Delta_0/\epsilon_f$, 
the window within which the 
maximum can be observed is narrowly 
centered around the nodal direction, 
as expected.
Finally, Fig.4 demonstrates the linear, in temperature, behavior predicted 
here for $T\ll T_c$, which is in a sharp contrast to the exponential 
decay of the Fermi function, and agrees remarkably well with
the result given in  Eq.(\ref{slope}).

The maximum of $\alpha_s$ is at high
temperature where the electronic contribution to the attenuation
is difficult to measure. While our results are qualitatively
the same for a different geometry of the Fermi surface, 
such as a tight-binding, the position of the maximum 
may be shifted; we also note that
for a 
tight-binding Fermi surface close to half-filling the peak does not exist:
large flat regions of the Fermi surface away from the nodes contribute to
the attenuation in the near-nodal direction above $T_c$, and
in the superconducting state
the loss of phase space in the gapped regions
cannot be compensated for by the
increase in scattering near the nodes.
Moderate impurity scattering
is expected to ``average'' the attenuation over a small range of angles. 
As seen from Fig.2, this can change the slope near the node, 
but has little effect on  the slope away form the node, and this result
should be robust with respect to scattering by
dilute impurity centers. Therefore measurement
of the low temperature slope
 provides a direct measurement of the parameter $v_2/v_f$, which determines
the low temperature behavior of the cuprates. 
We also note that we expect the general considerations of this work, 
including the breaking of particle-hole symmetry illustrated in Fig. 1b,
to be important 
for other transport coefficients, such as Hall effect, 
although a different and detailed analysis is needed there.

IV is grateful to L. Taillefer for important discussions. 
This research has been supported in part by NSERC of Canada (EJN and JPC), 
and CIAR (JPC). EJN is a Cottrell Scholar of Research Corporation.
\begin{figure}
\epsfxsize=3.5 in
\epsfbox{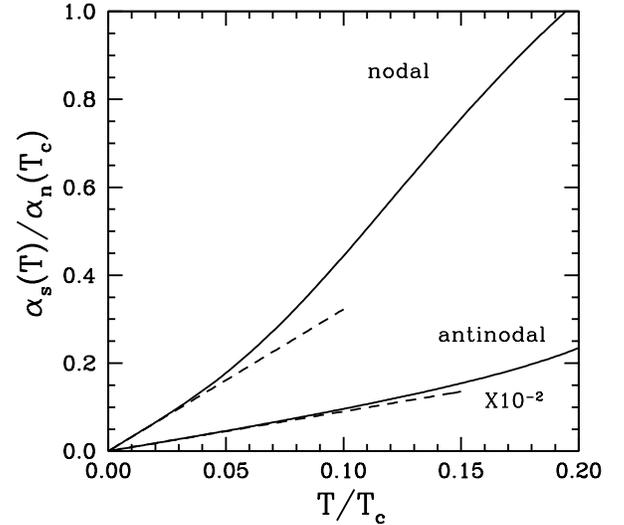}
\narrowtext
\caption{Low temperature dependence of the attenuation coefficient in the 
near-nodal ($\theta=0.78$) and anti-nodal ($\theta=1.57$) 
directions for  
$\Delta_0/\epsilon_f=0.1$. Dashed line: analytic result from Eq.(\ref{slope}).}
\end{figure}

\end{multicols}

\begin{references}

\bibitem{somereview} J. F. Annett, N. Goldenfeld, and S. R. Renn, in
{\it Physical Properties of High Temperature Superconductors II}, ed. 
by D. M. Ginzberg (World Scientific, Singapore, 1990), and references therein.

\bibitem{lee1} P. A. Lee, Phys. Rev. Lett. {\bf 71}, 1887 (1993).

\bibitem{lee} P. A. Lee and X.-G. Wen, Phys. Rev. Lett. {\bf 78}, 4111 (1997).

\bibitem{bonn1} D. A. Bonn {\it et al.}, Czech. J. Phys. {\bf 46}, S6, 3195
 (1996).

\bibitem{may} M. Chiao {\it et al.}, unpublished.

\bibitem{bhatt} S. Bhattacharya {\it et al.}, Phys. Rev. B {\bf 37}, 
5901 (1988).
\bibitem{ong} J. M. Harris {\it et al.}, J. Low Temp. Phys.
{\bf 105}, 877 (1996).

\bibitem{maki} H. Won and K. Maki, Phys. Rev. B {\bf 49}, 1397 (1994).

\bibitem{moreno} J. Moreno and P. Coleman, Phys. Rev. B {\bf 53}, R2995 (1996).

\bibitem{bonn2} A. Erb, E. Walker, R. Fl\"ukiger, 
	Physica C {\bf 258}, 9 (1996); R. Liang, D. A. Bonn, W. N. Hardy,
	Physica C {\bf 304}, 105 (1998).

\bibitem{bonn3} A. Hosseini {\it et al.}, unpublished

\bibitem{morse} R. W. Morse and H. V. Bohm, Phys. Rev. {\bf 108},
	1094 (1957).


\bibitem{pokr} V. L. Pokrovskii, Sov. Phys.-- JETP {\bf 13}, 628 (1961).

\bibitem{kostur} V. N. Kostur, J. K. Bhattacharjee, and R. A. Ferrell,
JETP Lett. {\bf 61}, 560 (1995).

\bibitem{swihart} T. Wolenski and J. C. Swihart, Physica C {\bf 253},
	266 (1995).

\bibitem{sue} S. N. Coppersmith and R. A. Klemm, Phys. Rev. Lett. {\bf 56},
	1870 (1986).
\bibitem{schr} J. R. Schrieffer, 
{\it Theory of Superconductivity} (Addison-Wesley, 1964).


\end{references}
\end{document}